\documentclass[10pt,twocolumn,letterpaper]{article}
\usepackage{iccv}
\usepackage{times}
\usepackage{epsfig}
\usepackage{graphicx}
\usepackage{amsmath}
\usepackage{amssymb}

\graphicspath{{./fig/}}
\DeclareGraphicsExtensions{.pdf,.jpeg,.png,.eps}
\usepackage{subfig}
\usepackage{multirow}
\usepackage{booktabs}
\usepackage[ruled, linesnumbered]{algorithm2e} 
\usepackage{diagbox}
\usepackage{makecell,rotating}

\usepackage[breaklinks=true,bookmarks=false]{hyperref}

\iccvfinalcopy 

\def\iccvPaperID{****} 

\ificcvfinal\fi

\begin{document}

\title{MFAGAN: A Compression Framework for Memory-Efficient On-Device
Super-Resolution GAN}

\author{Wenlong Cheng$^\#$\\
City University of Hong Kong\\
\and
Mingbo Zhao$^\#$\\
Donghua University\\
\and
Zhiling Ye\\
Tencent Computer System Co., Ltd.\\
\and
Shuhang Gu$^*$\\
The University of Sydney\\
}

\maketitle
\ificcvfinal\thispagestyle{empty}\fi

\begin{abstract}
Generative adversarial networks (GANs) have promoted remarkable advances in single-image super-resolution (SR) by recovering photo-realistic images. However, high memory consumption of GAN-based SR (usually generators) causes performance degradation and more energy consumption, hindering the deployment of GAN-based SR into resource-constricted mobile devices. In this paper, we propose a novel compression framework \textbf{M}ulti-scale \textbf{F}eature \textbf{A}ggregation Net based \textbf{GAN} (MFAGAN) for reducing the memory access cost of the generator. First, to overcome the memory explosion of dense connections, we utilize a memory-efficient multi-scale feature aggregation net as the generator. Second, for faster and more stable training, our method introduces the PatchGAN discriminator. Third, to balance the student discriminator and the compressed generator, we distill both the generator and the discriminator. Finally, we perform a hardware-aware neural architecture search (NAS) to find a specialized SubGenerator for the target mobile phone. Benefiting from these improvements, the proposed MFAGAN achieves up to \textbf{8.3}$\times$ memory saving and \textbf{42.9}$\times$ computation reduction, with only minor visual quality degradation, compared with ESRGAN. Empirical studies also show $\sim$\textbf{70} milliseconds latency on Qualcomm Snapdragon 865 chipset.   
\end{abstract}

\begin{figure}[!htp]
\centering
\hspace{-0.15in}
\begin{minipage}[b]{0.5\linewidth}
\includegraphics[width=0.83\linewidth]{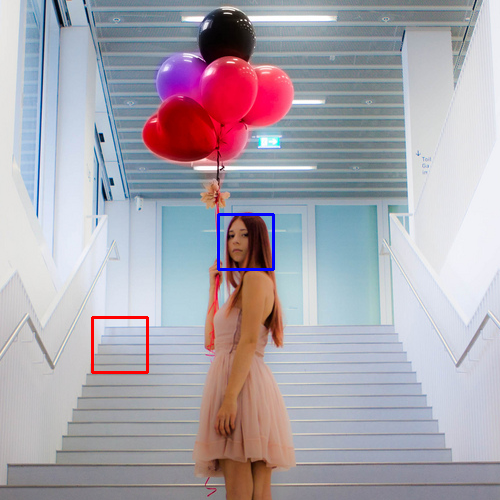}
\end{minipage}
\hspace{-0.2in}
\begin{minipage}[b]{0.5\linewidth}
\centerline{\subfloat{\includegraphics[width=0.36\linewidth]{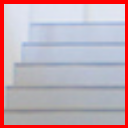}}\hspace{-2pt}\ \subfloat{\includegraphics[width=0.36\linewidth]{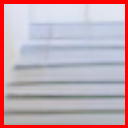}}\hspace{-2pt}\ \subfloat{\includegraphics[width=0.36\linewidth]{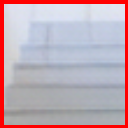}}}
\vspace{-0.14in}
\centerline{\subfloat{\includegraphics[width=0.36\linewidth]{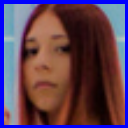}}\hspace{-2pt}\ \subfloat{\includegraphics[width=0.36\linewidth]{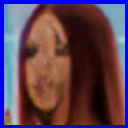}}\hspace{-2pt}\ \subfloat{\includegraphics[width=0.36\linewidth]{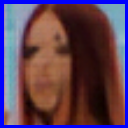}}}
\renewcommand\arraystretch{0.5}
\begin{tabular}{ccc}
\hspace{-14pt}\tiny HR& \tiny ESRGAN& \tiny MFAGAN\\
\hspace{-10pt}\tiny PSNR$\uparrow$ $\vert$ LPIPS$\downarrow$ &\tiny 31.13 $\vert$ 0.0632 &\tiny 32.59 $\vert$ 0.0514\\
\hspace{-13pt}\tiny $\#$FLOPs $\vert$ Memory &\tiny \hspace{-1pt}291G $\vert$  4.29G &\tiny 8.41G $\vert$ 0.551G
\end{tabular}
\end{minipage}
\vspace{-5pt}
\caption{4$\times$ SR results for the image ’img 009’ in Urban100. Our proposed MFAGAN reduces the memory usage of ESRAGN by 8.3$\times$ while preserving sharp edges and rich textures. \textbf{(Zoom in for best view)}}
\vspace{-15pt}
\label{f1}
\end{figure}

\section{Introduction}
Single image super-resolution (SR) is a fundamental low-level vision task, which aims to reconstruct a high-resolution (HR) image from a degraded low-resolution (LR) input. In recent years, convolutional neural networks (CNNs) based approaches \cite{dong2014learning,dong2016accelerating,shi2016real,zhang2018residual,kim2016deeply,ahn2018fast,zhang2018image,anwar2020densely} have achieved high fidelity in terms of peak signal-to-noise ratio (PSNR)  \cite{timofte2018ntire}. Nevertheless, these PSNR-oriented methods tend to produce blurry output without enough high-frequency information. More recently, an emerging direction is to resolve the ill-posed SR problem by using generative adversarial networks (GANs) \cite{goodfellow2014generative}. State-of-the-art GAN-based perceptual-driven models such as SRGAN \cite{ledig2017photo}, ESRGAN \cite{wang2018esrgan} can generate photorealistic images with more natural textures and sharper edges. With the popularity of mobile devices, there is a growing on-device demand for GAN-based SR applications.

However, the aforementioned GAN-based SR applications are extremely memory intensive and energy overhead, making it impractical to deploy GAN-based SR generators into resource-limited mobile devices. On the one hand, mobile devices are memory-constrained. For instance, a Snapdragon 865 GPU has 2800MB memory bandwidth, while ESRGAN would cost over 4000MB memory consumption to process a 256 $\times$ 256 image. Furthermore, the total memory bandwidth is shared by various on-device applications and the operating system. The peak bandwidth consumed by a single application may only be allocated 20–40\% \cite{lee2019device} of the total memory bandwidth. The overall performance of GAN-based SR is bound by the limited memory bandwidth and lives under the slanted part of the roofline \cite{lopes2017exploring}. On the other hand, mobile devices are energy-constrained. The energy consumption of GAN-based SR mainly comes from memory access cost and computation. With a 45 nm technology, the 32b coefficients in off-chip DRAM cost 640 pJ, 128$\times$ larger than the consumption of 32b coefficients in on-chip SRAM (3.7 pJ) \cite{han2016eie}. The heavy memory cost of GAN-based SR severely hinder the mobile deployment or at least creates performance degradation.

During the past few years, tremendous model compression techniques \cite{han2015deep} have been proposed to speed-up the inference and reduce the model parameters and GAN compression is recently a hot topic in this research area \cite{li2020gan,fu2020autogan,wang2020gan}. For example, Aguinaldo et al. \cite{aguinaldo2019compressing}, and Chen et al. \cite{chen2020distilling} first exploited different knowledge distillation modalities for CycleGAN compression. Gong et al. \cite{gong2019autogan}, Shu et al. \cite{shu2019co}, and Chu et al. \cite{chu2019fast} employed reinforcement learning or co-evolutionary learning-based search to accelerate GAN. However, these methods cannot be directly extended to compress GAN-based SR due to the following reasons: first, the minimax training of GAN is notoriously unstable and prone to collapse. This will greatly result in non-trivial solutions for some complex tasks such as GAN-based SR compression; second, memory efficiency is a critical issue when deploying the memory overhead models on the mobile phone. However, none of above compression methods have consider this point. Therefore, how to develop a memory efficiency and effective generator for GAN-based SR, is an urgent problem. 

In this work, in order to develop a hardware-aware on-device GAN-based SR network, we propose a new network, namely, \textbf{M}ulti-scale \textbf{F}eature \textbf{A}ggregation Net based \textbf{GAN} (MFAGAN) for memory efficient compression. In detail, we first propose a novel generator architecture by designing the multi-scale feature aggregation modules (MFAMs), which is memory-efficient and of sufficient expressive ability. Besides, we introduce the light-weight PatchGAN discriminator to overcome artifacts and facilitate MFAGAN training. Meriting from the above structures, we propose a two-stage compression method: we first utilize knowledge distillation both for the generator and discriminator. This is to balance the student generator and student discriminator for maintaining the stabilized training of the compressed generator so that the non-trivial solution can be achieved; we then apply NAS channel pruning method to further reduce memory usage. Finally, we utilize the hardware-aware evolutionary search for specializing SubGenerator on the target mobile phone.

The main contributions can be shown as follows: 1) we start from a new and more useful point of view to design a small-size SR method by firstly considering the memory efficiency. This is of great practice in real-world applications; 2) we develop a new two-stage approach to achieve effective compression, where we first distill G/D to achieve stabilized non-trivial solution of the compressed generator and then apply NAS to reduce memory usage; 3) Extensive experiments validate the efficiency of our method. MFAGAN achieves up to 8.3$\times$ memory reduction, 42.9$\times$ computing efficiency over ESRGAN with only minor loss in PSNR and Learned Perceptual Image Patch Similarity (LPIPS) \cite{timofte2018ntire}. Finally, we deploy our MFAGAN on OPPO Find X2 and demonstrate $\sim$70 milliseconds latency on Snapdragon 865 GPU.



\section{Related Work}
In this section, we review previous works about GAN-based super-resolution networks, efficient super-resolution networks, and GAN compression techniques, which are the most relevant to our work.

\subsection{GAN-based Super-Resolution Networks}
Recently, a bunch of SR works paid more attention to visual effects. With the rapid development of perceptual-driven SR algorithms, Generative Adversarial Network (GAN)-based methods often achieved state-of-the-art visual performance. Johnson et al.\cite{johnson2016perceptual} adopted the perceptual loss to enhance the visual quality while Ledig et al.\cite{ledig2017photo} firstly employed the adversarial loss to generate more realistic images. Besides, Sajjadi et al.\cite{sajjadi2017enhancenet} explored the texture matching loss to reduce visually unpleasant artifacts. Based on these works, Wang et al. \cite{wang2018esrgan} enhanced the SRGAN by employing Residual-in-Residual Dense Block (RRDB) and the relativistic discriminator, won the champion of PIRM2018-SR challenge.
Furthermore, the LPIPS metric was introduced to measure the perceptual similarity. Lately, Zhang et al.\cite{zhang2019ranksrgan} proposed a novel rank-content loss to optimize the perceptual quality, which achieved state-of-the-art results in perceptual metrics. Despite their performance boost, large GAN generators' growing complexity conflicts with the demands of mobile deployments, calling for GAN compression techniques. In this paper, we focus on memory-efficient GAN-based SR for mobile applications.

\subsection{Efficient Super-Resolution Networks}
In recent years, a series of efficient networks with parameters in the range of 10M have been proposed for the efficient SR task \cite{zhang2020aim}. We call these kinds of networks efficient super-resolution networks. They can be approximately divided into two classes: handcrafted architectures and model compression-based methods. A surge of handcrafted architectures have been designed for the efficient SR task, ranging from post-upsampling operators \cite{dong2016accelerating}, group convolutions \cite{hui2018fast}, residual blocks \cite{zhang2018residual}, recursive structures \cite{kim2016deeply}, cascaded architectures \cite{ahn2018fast}, inverse sub-pixel convolution \cite{shi2016real}, attention mechanisms \cite{zhang2018image}, to information multi-distillation block (IMDB) \cite{hui2019lightweight}. Besides, model compression techniques such as knowledge distillation, channel pruning, and binary quantization have also been used to speed up the SR networks. Specifically, RFDN \cite{liu2020residual} applied channel pruning along with residual feature aggregation module to improve the IMDB efficiency, which is the winner solution of the AIM 2020 Challenge on Efficient Super-Resolution \cite{zhang2020aim}. However, these methods aim to maximize PSNR between SR and HR, which tend to generate blurry results without high-frequency details. Efficient super-resolution networks cannot be directly used as the generator of GAN-based SR.

\subsection{GAN Compression Techniques}
In the past few years, GAN has achieved prevailing success in many generation and translation tasks. However, the growing memory complexity and computation cost of GANs conflict with the demands of mobile deployments. It is hard to apply existing compression techniques, owing to the minimax training of GANs is notoriously unstable and prone to collapse. Several methods exploited knowledge distillation for compressing the image translation models. Aguinaldo et al. \cite{aguinaldo2019compressing} first introduced knowledge distillation for unconditional GANs compression. Chen et al. \cite{chen2020distilling} proposed to train an efficient generator by knowledge distillation over the architectures of both the generator and the discriminator. A few works have also attempted to incorporate neural architecture search (NAS) \cite{zoph2016neural} with GANs. Gong et al. \cite{gong2019autogan} utilized reinforcement learning to search for an efficient generator with a fixed discriminator, limiting the algorithm to discover an optimal generator since the balance between these two players needs to be considered. Shu et al. \cite{shu2019co} later replaced the reinforcement learning by co-evolutionary pruning to accelerate CycleGAN, which relied on the cycle consistency loss. Chu et al. \cite{chu2019fast} leveraged an elastic search tactic at both micro and macro space to solve a multi-objective problem for SR. Combining multiple different compression techniques, such as weight sharing, channel pruning, knowledge distillation, and quantization, has been significantly outperformed approaches using single compression techniques. Li et al. \cite{li2020gan} presented a compression framework for conditional GANs via intermediate feature distillation and automated channel reduction in a ``once-for-all'' manner. Fu et al. \cite{fu2020autogan} performed computational resources constrained differential neural architecture search via the guidance of knowledge distillation. Wang et al. \cite{wang2020gan} combined three compression techniques: model distillation, channel pruning, and quantization, together with the minimax objective, into one unified optimization to form an end-to-end optimization framework.

However, most of the above methods are not customized for GAN-based SR. Besides, they have not yet addressed the memory intensive problem. The memory budget should still be met for efficient network structure design. Moreover, the mutual balance of the compressed generator and discriminator needs to be considered, which is crucial for stabilizing the GANs training process. Our work is to solve the above problems.

\section{Proposed Approaches}
\subsection{Network Structure}
\begin{figure}
\centerline{\includegraphics[width=0.44\textwidth]{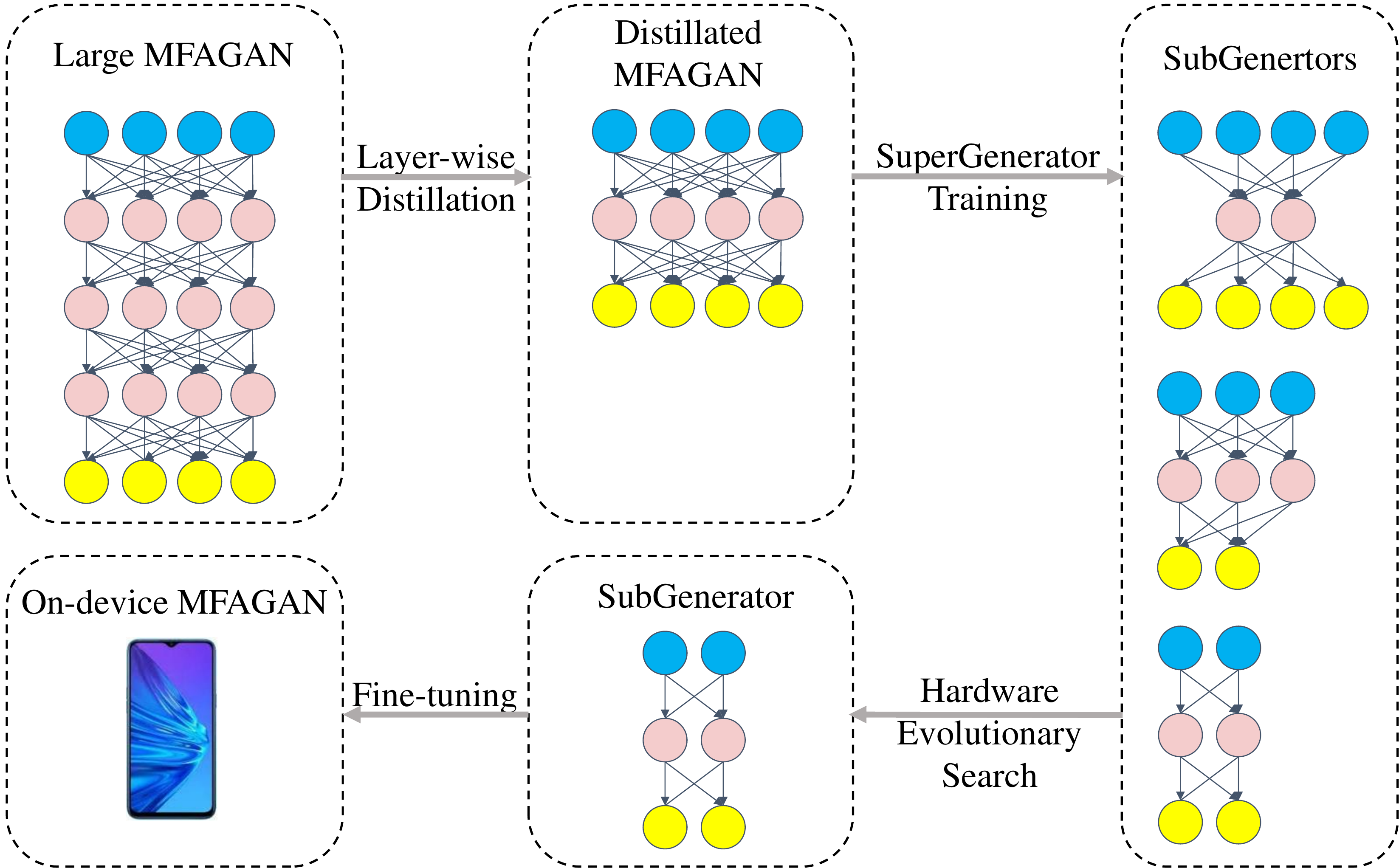}}\
\vspace{-15pt}
\caption{MFAGAN overview. (1) Construct the large MFAGAN (MFAGAN$\_$L) using the proposed Multi-Scale Feature Aggregation Network as a generator and the introduced PatchGAN discriminator. (2) Distill both the generator and discriminator in the MFAGAN$\_$L. (3) Train a weight-shared SuperGenerator, which comprises many SubGenerators of different channel numbers. (4) Perform a hardware-aware evolutionary search to find the satisfactory SubGenerator. (5) Fine-tune the searched SubGenerator with the previously distilled discriminator.}
\vspace{-15pt}
\label{f2}
\end{figure}

In our network, we first develop a novel generator architecture by designing the multi-scale feature aggregation modules (MFAMs). We then introduce the light-weight PatchGAN discriminator to overcome artifacts and facilitate MFAGAN training. Meriting from the above structures, we then distill both generator and discriminator to achieve non-trivial super-generator and to apply NAS channel pruning on it.  The overall proposed compression framework architecture is depicted in Figure \ref{f2}.

\subsection{Multi-Scale Feature Aggregation Network}
\begin{figure*}\hspace{0.2in}
\centerline{\includegraphics[width=0.9\textwidth]{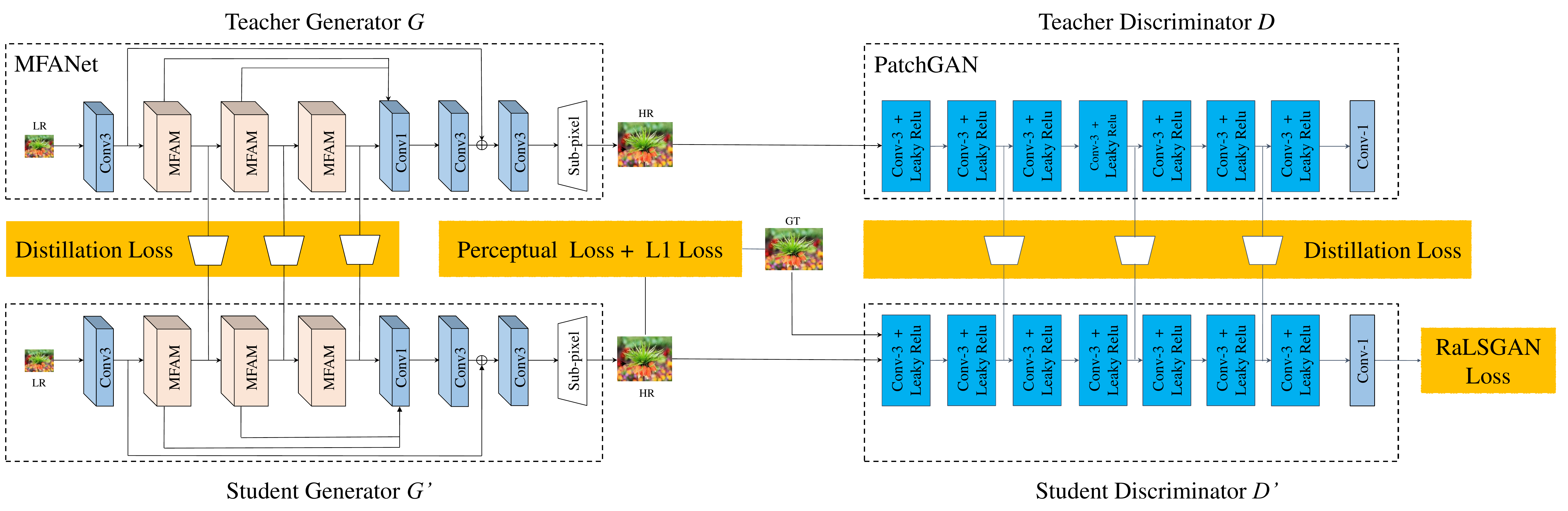}}
\vspace{-10pt}
\caption{Illustration of the overall objective.}

\vspace{-10pt}
\label{f5}
\end{figure*}

For the generator, inspired by ESRGAN \cite{wang2018esrgan}, VoVNet \cite{lee2019energy} and IMDN \cite{hui2019lightweight}, we design a memory-efficient Multi-Scale Feature Aggregation Network (MFANet) shown in Figure \ref{f5}. The proposed MFANet mainly contains four parts: the coarse feature extraction part, the multi-scale feature extraction part, the feature fusion part, and the reconstruction part. In particular, we use a $3\times3$ convolution as the coarse feature extraction part to generate features from the input LR image. The following is the multi-scale features extraction part, in which three Multi-scale Feature Aggregation Modules (MFAMs) are stacked in a chain manner to refine the extracted features gradually. Later we will provide a detailed description of MFAM. After extracting multi-scale features with a set of MFAMs, we further conduct global feature aggregation by a $1\times1$ convolution layer, which contributes to concatenate multi-scale features of different modules and reduces computation complexity.
Meanwhile, a global skip connection is applied between different scale features so that the features information can be fully exploited. Then, a $3\times3$ convolution layer is used to smooth the aggregated features. Finally, the HR images are generated by the image reconstruction part, which only consists of a $3\times3$ convolution and a sub-pixel operation.  
\begin{figure}\centerline{\subfloat[RFDB]{\includegraphics[width=0.33\linewidth]{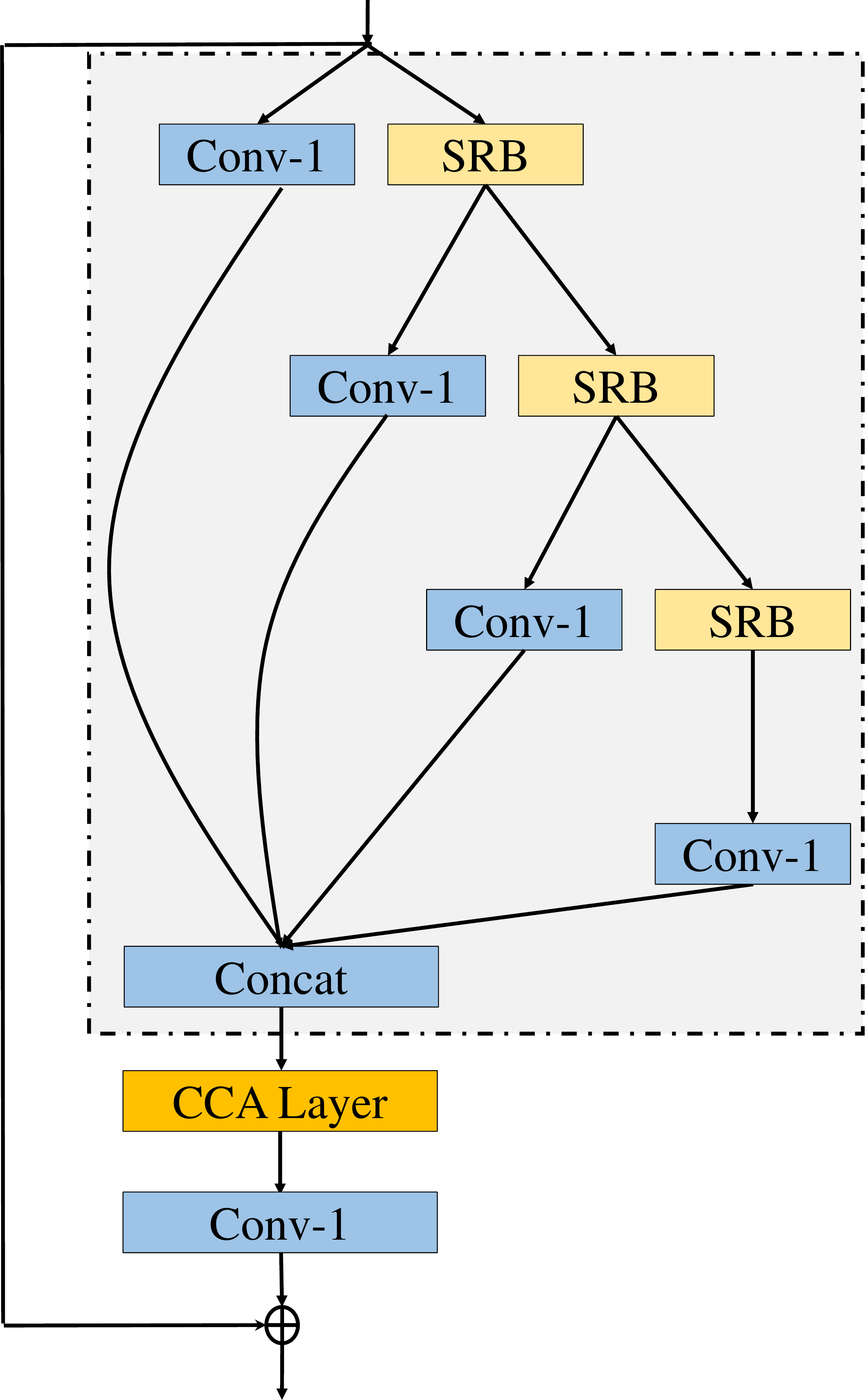}}\hspace{0.08in}
\subfloat[MFAM(ours)]{\includegraphics[width=0.33\linewidth]{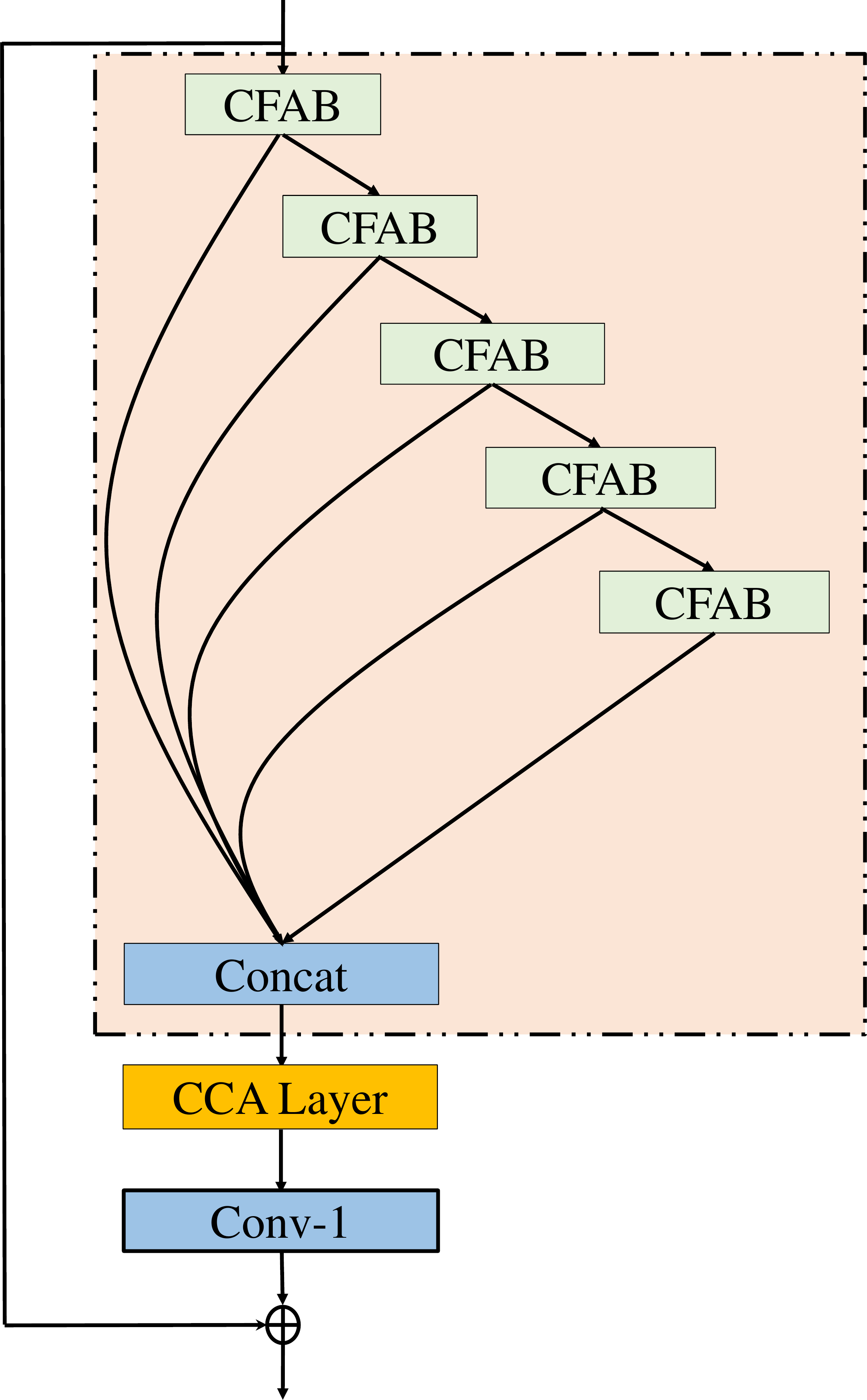}}\hspace{0.08in}
\subfloat[CFAB(ours)]{\includegraphics[width=0.33\linewidth]{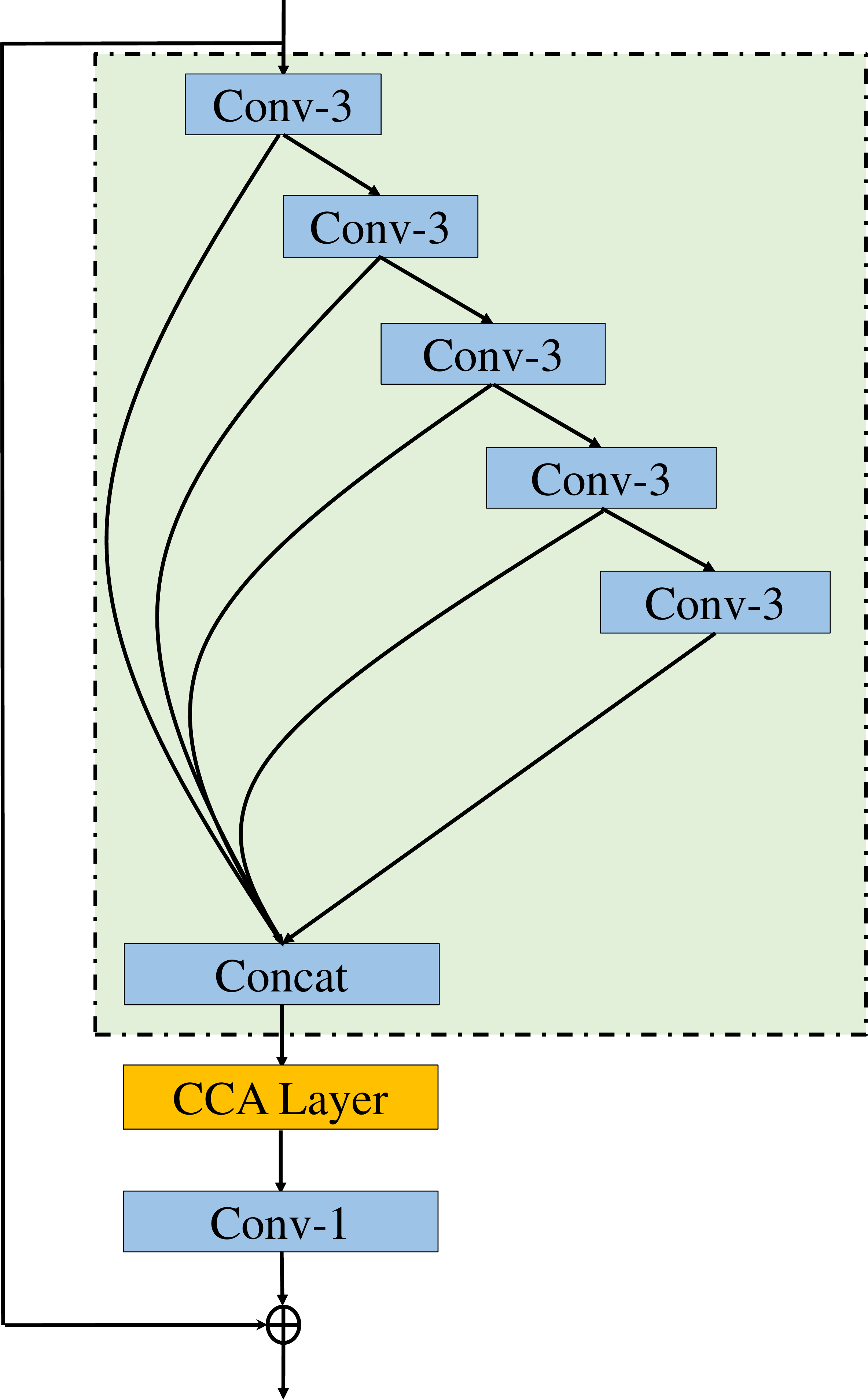}}}
\caption{Illustration of three structures. (a) RFDB: the residual feature distillation block. (b) MFAM: the multi-scale feature aggregation module. (c) CFAB: the concatenative feature aggregation block in MFAM.}
\vspace{-15pt}
\label{f4}
\end{figure}

Modifying from the RFDN \cite{liu2020residual}, as illustrated Figure \ref {f4} (a), we propose the Multi-scale Feature Aggregation Module (MFAM), as shown in Figure \ref{f4} (b), which is more expressive and memory-efficient than the RFDB. In RFDB, we can see that the channel reduction is conducted by a $1\times1$ convolution on the left, which compresses feature channels at a fixed ratio. Although the $1\times1$ convolution decreases the number of parameters, it increases the overall memory complexity. Therefore, the three $1\times1$ convolutions are removed, since we find that the NAS method is more efficient for channel reduction without introducing extra computation. Moreover, we also introduce the Concatenative Feature Aggregation Block (CFAB), as displayed in Figure \ref{f4} (c), to replace SRB in RFDB \cite{liu2020residual}. CFAB consists of five $3\times3$ convolutions, a local feature aggregation layer, a CCA layer \cite{hui2019lightweight}, a $1\times1$ convolution layer, and a local skip connection. CFAB can conserve feature representations of multiple receptive fields as well as preserve original information. This schema provides diverse information for recovering high-resolution details. In summary, the main goal of MFAM is to reduce memory overhead and enhance expressive ability.

\subsection{PatchGAN Discriminator}
Besides the memory-efficient generator, the ESRGAN discriminator is replaced with the PatchGAN discriminator \cite{demir2018patch}. We utilize a 7-layer fully convolutional discriminator. Each convolutional layer is followed by a leaky ReLU. To avoid unpleasant artifacts, all BN layers are removed. Compared with the original discriminator, the PatchGAN discriminator has fewer parameters. Another advantage is that it only models local patches instead of the whole image, making the MFAGAN$\_$L training faster and more stable.  
\vspace{-5pt}

\subsection{Knowledge Distillation}
\textbf{Layer-wise knowledge from the teacher generator.} Several attempts have been made to compress GANs generator with knowledge distillation in image translation. In this work, we match the teacher generator's intermediate representations, as the layer-wise knowledge distillation work in \cite{romero2014fitnets}. In particular, we first train a teacher generator $G$ until convergence and then conduct layer-wise knowledge transfer from $G$ to the student generator $G'$. The goal of distillation is that the feature maps of each layer in $G'$ should be as close as possible to those of $G$. Feature maps of the 1st, 2nd and 3rd MFAMs of $G$ denoted as $t_1$, $t_2$ and $t_3$ respectively. The corresponding levels of feature maps in $G'$ are the outputs of the 1st, 2nd, and 3rd MFAMs, denoting as $s_1$, $s_2$, and $s_3$, respectively. To address the above issue, we use the information in $t_1$, $t_2$ and $t_3$ to guide the information $s_1$, $s_2$ and $s_3$ during the training of $G'$. Subsequently, a student generator of fewer channels is trained by inheriting the low-level and high-level information from the original heavy teacher generator. 

\textbf{Layer-wise knowledge from the teacher discriminator.} Although we aim to compress the generator, a discriminator stores useful knowledge of a learned GAN to guide the training generator \cite{chen2020distilling}. Using heavy discriminator, $D$ sometimes leads to severe training instability and image quality degradation after the generator $G$ is compressed. It is necessary to distill the teacher discriminator $D$ to assist the training of the compressed generator $G'$. In this work, we adopt the PatchGAN discriminator architecture to distill layer-wise knowledge from teacher discriminator $D$ to student discriminator $D'$. Concretely, we extract corresponding levels of feature maps using the outputs of every two convolution layers, denoting as $t_2$, $t_4$, $t_6$, respectively. After that, we use the information in $t_2$, $t_4$, $t_6$ to guide the information $s_2$, $s_4$, $s_6$ during training of student discriminator. We jointly optimize $G'$ and $D'$ to minimize the distillation loss $L_{Distill\_G}$ and $L_{Distill\_D}$.

Mode collapse frequently occurs when the generator and discriminator are imbalanced. We adopt the layer-wise knowledge distillation on both the teacher generator and teacher discriminator. Hence, the student generator and discriminator are better matched. Such guidance from the teacher networks provides stable supervision in the early training phase of compressed networks. It is easy to implement, and mode collapse has never been experienced with our knowledge distillation schema.
\subsection{Objective Function}
\textbf{Overall objective.} There are five loss functions applied to training,  depicted in Figure \ref{f5}. The overall objective for our MFAGAN is the weighted sum of loss terms, written as follows:
\vspace{-0.1in}
\begin{equation}
\begin{aligned}
L = {\lambda _1}{L_{recon}}& + {\lambda _2}{L_{Distill\_G}} + {\lambda _3}{L_{Distill\_D}}\\
&  + {\lambda _4}{L_{percep}} + {\lambda _5}{L_{G}},
\end{aligned}
\vspace{-0.08in}
\label{e6}
\end{equation}
where $\lambda_1$, $\lambda_2$, $\lambda_3$, $\lambda_4$, and $\lambda_5$ are the trade-off hyper-parameters to balance different objectives.

\textbf{Reconstruction loss.} Here, we apply the reconstruction loss, specifically the Mean Absolute Error (MAE) loss, to enhance the fidelity of the recovered images. Reconstruction loss is trained to optimize the $L1$ distance between the recovered images and ground-truths:
\vspace{-0.08in}
\begin{equation}
{L_{recon}} = \frac{1}{N}\sum\limits_{i = 1}^N {{{\left\| {R\left( {x_i^{lr}} \right) - x_i^{gt}} \right\|}_{\rm{1}}}},
\label{e1}\vspace{-0.05in}
\end{equation}
where $x_{i}^{lr}$, $x_{i}^{gt}$ denote the $i$-th LR image patch and the corresponding HR. $N$ is the total number of training samples. $R(\cdot)$ represents the super-resolved output by MFAGAN.

\textbf{Layer-wise knowledge distillation loss.} 
We introduce a layer-wise knowledge distillation to extract the intermediate feature maps of the teacher generator. The intermediate feature maps contain richer information and allow the student generator to acquire low-level and high-level information from the teacher generator. The generator distillation objective can be formalized as: 
\label{e1}\vspace{-0.05in}
\begin{equation}
{L_{Distill\_G}}{\rm{ = }}\frac{1}{n}\sum\limits_{i = 1}^n {{{\left\|{{G_{i}}\left( {{x}} \right) - { G'_{i}}\left( {{x}} \right)} \right\|}_2}},
\label{e2}
\vspace{-0.05in}
\end{equation}
where ${G_{i}}({x})$ and ${G'_{i}}({x})$ are the intermediate feature maps of the $i$-th chosen layer in the teacher and student generator.
Besides the generator, the discriminator stores useful knowledge of a learned GAN-based SR. It is useful to distill the teacher discriminator to stabilize the compressed generator training. The discriminator distillation loss function can be defined as:
\label{e1}\vspace{-0.05in}
\begin{equation}
{L_{Distill\_D}}{\rm{ = }}\frac{1}{m}\sum\limits_{i = 1}^m {{{\left\|{{D_{i}}\left( {{y}} \right) - { D'_{i}}\left( {{y}} \right)} \right\|}_2}},
\label{e3}
\vspace{-0.05in}
\end{equation}
where ${D_{i}}({y})$ and ${D'_{i}}({y})$ are the feature maps of the $i$-th chosen layer in the teacher and student discriminator.

\textbf{Perceptual loss.} 
In \cite{johnson2016perceptual} Johnson et al. proposed the perceptual loss to improve the visual effect of low-frequency features such as edges. Instead of computing distances in image pixel space, the images are first mapped into feature space by a pre-trained VGG19 network, denoted as $\phi$, and then compute the Mean Square Error (MSE) on their feature maps as follows:
\vspace{-0.06in}
\begin{equation}
{L_{percep}} = \sum\limits_{i = 1} {{{\left\| {{\phi}\left({ \hat{y_i}} \right) - {\phi}\left({{y_i}} \right)} \right\|}_2}},
\label{e4}
\vspace{-0.06in}
\end{equation}
where $\phi ({{\hat{y}}_{i}})$ and $\phi ({{y}_{i}})$ represent the feature maps of HR ground truth and the SR, respectively. All the feature maps are obtained by the fourth convolutional layer before the fifth max-pooling layer within the VGG19 network. 

\textbf{Adversarial loss.} Following common practice, we apply adversarial loss \cite{wang2018esrgan} to enhance the texture details of the generator generated image to make it more realistic. Adversarial training a standard minimax optimization, and the discriminator $D$ is trained to distinguish between real images and the output of $G$. The adversarial loss ${{L}_{G}}$ is described as:
\begin{equation}
{L_{G}} =  - \log (D(G({I^{LR}})).
\label{e5}
\end{equation}
where $I^{LR}$ is the LR image, $D(G(I^{LR}))$ means the probability of the discriminator over all training samples.

\subsection{Hardware-aware NAS based Channel Pruning}
\textbf{SuperGenerator training with fine-grained channels.} To address the fine-grained channel pruning problem, we first build a SuperGenerator that comprises all candidate SubGenerators. Concretely, we use MFANet as the backbone to build a ``once-for-al'' \cite{cai2019once} network that comprises many SubGenerators of different channel numbers (i.e., 48, 32, 24), in which the full-width model is the SuperGenerator. The combined search space contains about $3^8 = 6581$ different SubGenerators, in which every SubGenerator in the search space is a part of the SuperGenerator. In practice, the SuperGenerator only needs to be trained for the same steps as a baseline SR model, which is fast and low-cost. We thus use the most important channels of the SuperGenerator to initialize the SubGenerators. All SubGenerators share the front portion of corresponding layer weights in the SuperGenerator.

\textbf{Hardware-aware evolutionary search for specialized SubGenerator.} After SuperGenerator training, we adopt the evolutionary search to find the satisfactory SubGenerator, which satisfies the target hardware's latency constraints while optimizing the PSNR. We can first build a lookup table containing the latency for all possible operators on the target hardware, then the overall latency of a SubGenerator is predicted by summing up each operator's latency \cite{wu2019fbnet}. Therefore, we can approximate the latency of candidate SubGenerator by querying the lookup table. The PSNR of SubGenerators is evaluated on the validation set. Afterward, we conduct the evolutionary search to get a specialized SubGenerator. Since SubGenerators training has been decoupled from the architecture search, we do not need any training cost in the search stage. This hardware-aware NAS channel pruning enables us to design a specialized SubGenerator on the target hardware.

Finally, we can fine-tune the pruned SubGenerator with distilled discriminator to obtain the final model.

\section{Experimental Results}
\subsection{Experimental Setup}
\textbf{Datasets.} Following \cite{wang2018esrgan}, we adopt 800 HR images from the DIV2K dataset \cite{agustsson2017ntire} as the training set. To generate LR training patches, we down-sample the HR images using bicubic interpolation in MATLAB. We also augment the training data with the random crop, horizontal/vertical flips, and $90^{\circ}$ rotations.

\textbf{Evaluation metrics.} For evaluation, we test the compressed/searched model on four SR benchmark datasets, namely Set5 \cite{bevilacqua2012low}, Set14 \cite{zeyde2010single}, B100 \cite{timofte2014a}, and Urban100 \cite{huang2015single}. Inspired by the PIRM2018-SR Challenge \cite{timofte2018ntire}, we introduce PSNR and LPIPS on the Y channel of the transformed YCbCr space as the quality evaluation metrics. To measure the computation efficiency, we compare the widely used metrics - memory access cost, parameters, FLOPs (floating-point operations), and inference latency. Regarding the inference latency, we use the published codes of competitors to evaluate on a server with 4.2GHz Intel i7 CPU, 32GB RAM, and an Nvidia V100 GPU card.

\textbf{Implement details.} The training process is divided into four main stages. (1) Constructing the MFAGAN$\_$L model. We first train an MFANet with the $L_1$ loss, while the learning rate is $2 \times {10^{-4}}$ and 500K iterations. The MFANet initialized generator is then trained using the loss function in Eq.\eqref{e6} with $\lambda_1 = 1$, $\lambda_4 = 1$, $\lambda_5 = 10$, the learning rate is initialized to $1 \times {10^{-4}}$ and halved at $[5k, 10k]$ iterations. (2) Layer-wise distillation on both the generator and discriminator in the MFAGAN$\_$L. The student generator and student discriminator is trained with  $\lambda_1 = 1$, $\lambda_2 = 0.05$, $\lambda_3 = 0.05$, $\lambda_4 = 1$, $\lambda_5 = 10$. (3) Training the weight-shared SuperGenerator. SuperGenerator is trained with $\lambda_1 = 1$, $\lambda_4 = 1$, the learning rate is set to $1 \times {10^{-4}}$ and halved at $[200k, 400k, 600k]$ iterations. (4) Fine-tuning the searched generator with distilled discriminator about 10K iterations. For all experiments, we use the Adam \cite{kingma2014adam} optimization method with $\beta_1 = 0.5$, $\beta_2 = 0.999$, $\epsilon = {10^{-8}}$ to train all of the models. The mini-batch size is set to 16. Our networks are implemented using the PyTorch framework on 8 NVIDIA V100 GPUs. The entire training process takes about 120 GPU hours.
\subsection{Model Complexity Analysis}
\begin{figure}
\centerline{\includegraphics[width=0.4\textwidth]{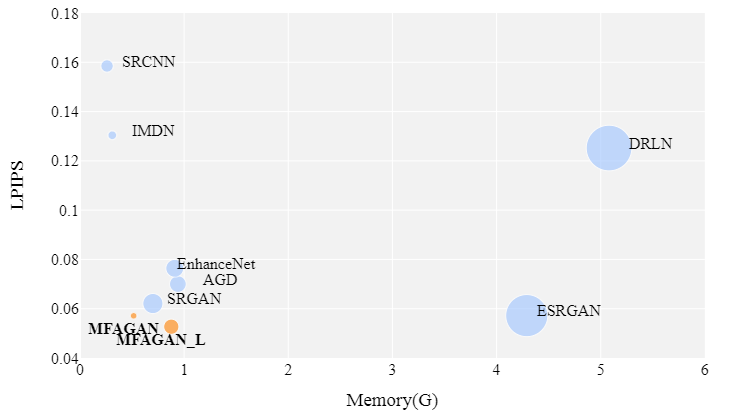}}
\vspace{-5pt}
\caption{The memory access cost vs. LPIPS on Set5 (4$\times$) dataset. The orange circles represent our proposed models. The circles' size represents the number of FLOPs, which are calculated on $512\times512$ HR image.}
\vspace{-10pt}
\label{f7}
\end{figure}
\vspace{-5pt}
To construct a memory-efficient SR model, the memory access cost of the network is vital. As discussed in previous sections, the proposed MFAGAN could significantly reduce memory consumption. Figure \ref{f7} depicts the comparisons about LPIPS vs. memory access cost and FLOPs on Set5 4$\times$ dataset. From Figure \ref{f7}, we can observe that the MFAGAN model with 3 MFAMs exhibits comparative or better performance and fewer memory usage than other state-of-the-art methods SRGAN \cite{ledig2017photo}, EhanceNet \cite{sajjadi2017enhancenet}, ESRGAN \cite{wang2018esrgan}, and AGD \cite{fu2020autogan}. MFAGAN is also superior over efficient models including SRCNN \cite{dong2014learning}, IMDN \cite{hui2019lightweight}. Compared with IMDN, our MFAGAN achieves better LPIPS with a slightly larger model. These results demonstrate that the proposed MFAGAN can correctly balance memory complexity and reconstruction performance.

\subsection{Ablation Study}
In this section, we conduct ablation experiments to investigate the contributions of each component in the proposed method. The overall comparison is illustrated in Table \ref{t1}, in which each column represents a model. A detailed discussion is provided as follows.

\begin{table}[htp]\footnotesize
\setlength{\tabcolsep}{1pt}
\caption{Ablation study: Memory-efficient architecture combined with knowledge distillation on both ${G}$ and ${D}$, and NAS channel pruning achieves the best performance on the Set5 dataset.}
\vspace{-5pt}
\newcommand{\tabincell}[2]{\begin{tabular}{@{}#1@{}}#2\end{tabular}}
\centering
\begin{tabular}{ccccccc}
\toprule
Options &\tabincell{c}{Baseline\\(ESRGAN)}&1st&2nd&3rd&4th&5th\\
\midrule
generator channels	    &64 &64 &48 & 48& 32& 32\\
discriminator channels	&64 &48 &32 & 48& 48& 32\\
\midrule
MFAGAN$\_$L&\multirow{5}{*}{}& $\checkmark$	& $\checkmark$ & $\checkmark$ & $\checkmark$ & $\checkmark$ \\	
\tabincell{c}{Distill $G$ and $D$\\}& & &$\checkmark$& & &$\checkmark$	\\
\tabincell{c}{Distill $G$\\}& & & &$\checkmark$& &    	\\
NAS pruning& & & & & $\checkmark$ & $\checkmark$ 	\\
\midrule
\tabincell{c}{PSNR$\uparrow$ \\LPIPS$\downarrow$}&
\tabincell{c}{30.45 \\ 0.0572}&	
\tabincell{c}{30.32 \\ 0.0527}&	
\tabincell{c}{30.65 \\ 0.0558}&	
\tabincell{c}{28.56 \\ 0.0878}&
\tabincell{c}{29.19 \\ 0.0672}&
\tabincell{c}{30.16 \\ 0.0571}	\\
Memory (G)	&4.29&	0.877&	0.657&	0.657&	0.515&	0.515	\\
\#Param. (MB)&	16.67&	1.56&	0.884&	0.884&	0.551&	0.551\\
\#FLOPs (G)	&291&	23.72&	13.45	&13.45&	8.41&	8.41\\
\bottomrule
\end{tabular}
\vspace{-10pt}
\label{t1}
\end{table}

\begin{table*}[htp]\footnotesize
\setlength{\tabcolsep}{5pt}
\caption{Quantitative comparison of our model with state-of-the-art perceptual-driven works on 4$\times$ SR task. In each row, \textcolor{red}{red}/\textcolor{blue}{blue} represents \textcolor{red}{best}/\textcolor{blue}{second}, respectively.}
\newcommand{\tabincell}[2]{\begin{tabular}{@{}#1@{}}#2\end{tabular}}
\vspace{-5pt}
\centering
\begin{tabular}{ccccccccc}
\toprule
Model&	
\tabincell{c}{Memory\\(G)}&	
\tabincell{c}{\#FLOPs\\(G)}&\tabincell{c}{\#Param.\\(MB)}&	
\tabincell{c}{Latency\\(ms)}&	\tabincell{c}{Set5\\PSNR$\uparrow$ LPIPS$\downarrow$}&
\tabincell{c}{Set14\\PSNR$\uparrow$ LPIPS$\downarrow$}&	
\tabincell{c}{B100\\PSNR$\uparrow$ LPIPS$\downarrow$}&	
\tabincell{c}{Urban100\\PSNR$\uparrow$ LPIPS$\downarrow$}\\
\midrule
SRGAN	&\textcolor{blue}{0.7}	&36.5&	1.55&	80.8&	
\tabincell{c}{29.40 $\vert$ 0.0621}&
\tabincell{c}{26.11 $\vert$ 0.1167}&
\tabincell{c}{25.17 $\vert$ 0.1333}&
\tabincell{c}{$-$}\\

EnhanceNet&	0.91&	30.2	&0.85&	29&		
\tabincell{c}{28.56 $\vert$ 0.0764}	&	
\tabincell{c}{25.77 $\vert$ 0.1295}	&	
\tabincell{c}{24.93 $\vert$ 0.1481}&		
\tabincell{c}{23.54 $\vert$ 0.1307}	\\

ESRGAN	&4.29	&291&	16.67&	169.4	&	
\tabincell{c}{\textcolor{red}{30.45} $\vert$ \textcolor{blue}{0.0572}}	&	
\tabincell{c}{26.28 $\vert$ \textcolor{red}{0.1055}}&		
\tabincell{c}{25.32 $\vert$ \textcolor{red}{0.1216}}&	
\tabincell{c}{\textcolor{blue}{24.36} $\vert$ \textcolor{red}{0.1000}} \\

\tabincell{c}{AGD}&	0.94&	\textcolor{blue}{27.1}&	\textcolor{red}{0.41}	&\textcolor{blue}{27.7}&		
\tabincell{c}{\textcolor{blue}{30.41} $\vert$ 0.0700} &		
\tabincell{c}{\textcolor{red}{27.27} $\vert$ 0.1247}&		
\tabincell{c}{\textcolor{red}{26.22} $\vert$ 0.1556}&	
\tabincell{c}{\textcolor{red}{24.73} $\vert$ 0.1329}\\

\tabincell{c} {\textbf{MFAGAN (ours)}}	&\textcolor{red}{0.52}&\textcolor{red}{8.41}&	\textcolor{blue}{0.55}&	\textcolor{red}{21.9}&		
\tabincell{c}{{30.16} $\vert$ \textcolor{red}{0.0571}}	&	
\tabincell{c}{\textcolor{blue}{26.69} $\vert$ \textcolor{blue}{0.1133}}&		\tabincell{c}{\textcolor{blue}{25.33} $\vert$ \textcolor{blue}{0.1332}} &	
\tabincell{c}{24.23 $\vert$ \textcolor{blue}{0.1132}}\\
\bottomrule
\end{tabular}
\label{t2}
\vspace{-10pt}
\end{table*}

\textbf{Effectiveness of memory-efficient architecture.} We first
analyze the advantage of MFANet based MFAGAN$\_$L. As shown in the 1st column, MFAGAN$\_$L has comparative PSNR and LPIPS results with the baseline model. While achieves 4.89$\times$ memory saving and 12.26$\times$ computation reduction. With our memory consumption and computation complexity largely reduced, the SR performance remains relatively stable. The perceptual SR task requires the model to be expressive enough to recover more realistic texture details. Our proposed multi-scale feature aggregation modules (MFAMs) is capable of aggregating multi-scale features to produce powerful feature representation.

\textbf{Effectiveness of layer-wise knowledge distillation on both ${G}$ and ${D}$.} We also investigate the effects of different distillation objectives on the MFAGAN$\_$L. Two distillation methods are proposed, including solely layer-wise distillation on teacher generator $G$, and knowledge distillation on both teacher generator $G$ and teacher discriminator $D$. Results are illustrated in the 2nd and 3rd columns of Table \ref{t1}, which show that the proposed distillation objective is useful for the MFAGAN$\_$L compression. As shown in the 2nd column, distilled MFAGAN$\_$L even has better PSNR result than MFAGAN$\_$L, with 1.33$\times$ memory saving and 1.76$\times$ computation reduction. While solely distillation on $G$ yields worse performance compared with MFAGAN$\_$L. As a teacher discriminator, $D$ stores useful information about teacher generator $G$. It can offer strong supervision to guide the student generator $G'$ to learn faster and better.

\textbf{Effectiveness of NAS channel pruning.} We further explore the role of NAS channel pruning. The results are shown in the 4th and 5th columns of Table \ref{t1}. Directly using NAS channel pruning to compress the MFAGAN$\_$L generator largely degrades the image SR performance. In contrast, knowledge distillation + NAS channel pruning achieves much better PSNR and LPIPS results, showing the necessity of jointly using channel pruning and knowledge distillation. Actually, the capacity gap between the SuperGenerator, i.e., MFAGAN-64, and the directly searched SubGenerator, i.e., MFAGAN-32, are too huge. As a result, the inherited weights from the SuperGenerator may be too recondite for the SubGenerator, in which case large ratio NAS channel pruning would have negative effects on the searched model. Applying the distilled MFAGAN-48 as SuperGenerator allows us to find a SubGenerator, which has a smaller gap between the SuperGenerator and hence makes learning easier. NAS channel pruning and knowledge distillation are complementary to each other, which guarantees MFAGAN achieve competitive results. 

\begin{figure}
\centerline{\includegraphics[width=0.4\textwidth]{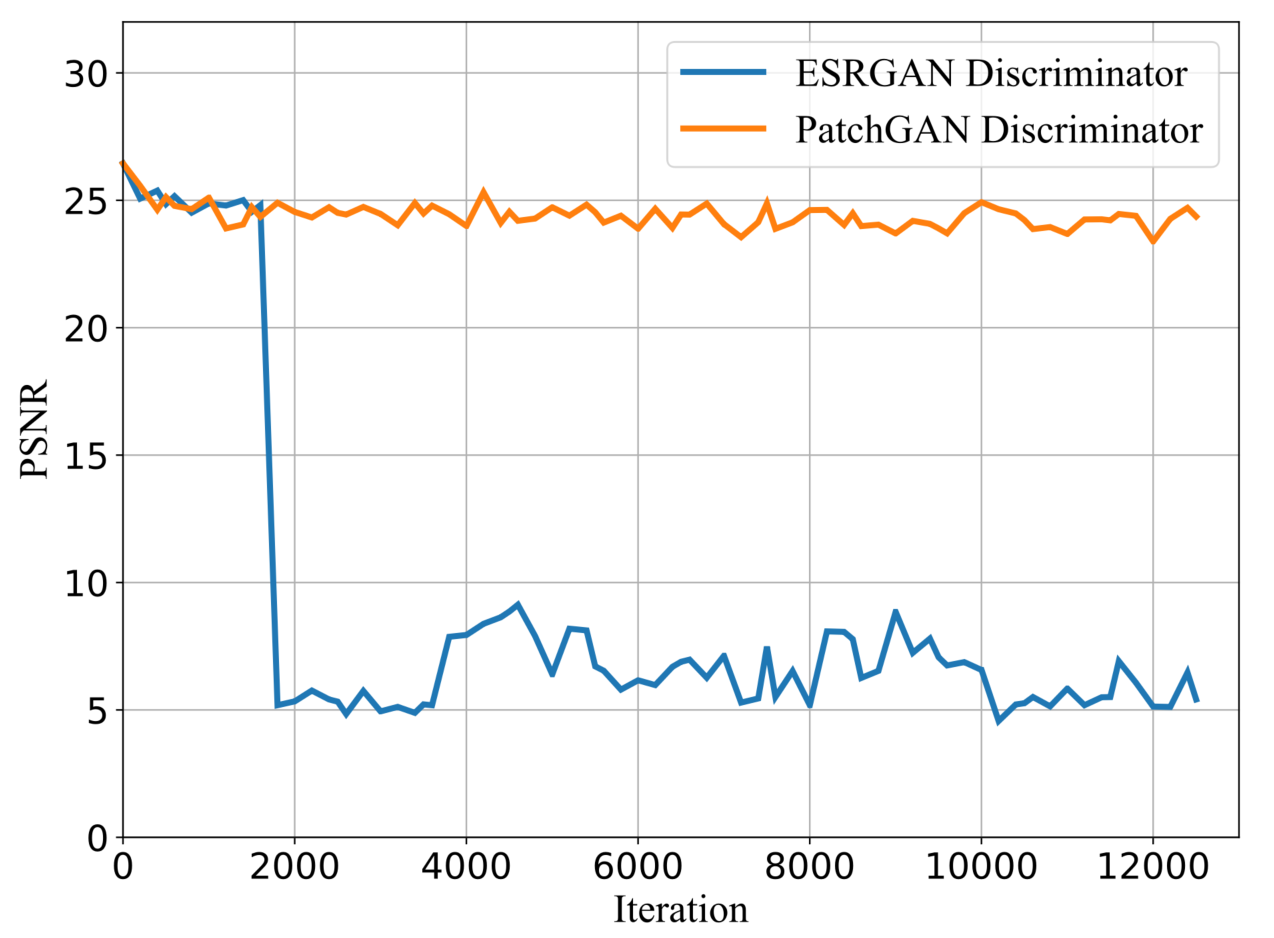}}
\caption{Ablation study of PatchGAN discriminator.}
\vspace{-8pt}
\label{f8}
\end{figure}

\textbf{Effectiveness of discriminator.} Finally, this section aims to evaluate the importance of the PatchGAN discriminator. To this end, we train the MFAGAN$\_$L with PatchGAN discriminator and ESRGAN discriminator on DIV2K, respectively. The convergence curves are visualized in Figure \ref{f8}. We can observe that PatchGAN discriminator achieves better results, verifying that: (1) vanilla ESRGAN discriminator is not matched with light-weight MFANet and has worse performance; (2) PatchGAN discriminator leads to significantly better PSNR result and stability. 

\vspace{5pt}
\vspace{-8pt}

\subsection{Comparison with State-of-the-Art Methods}
\textbf{Quantitative comparisons.} We report the quantitative comparisons of state-of-the-art perceptual-driven methods over the benchmark datasets in Table \ref{t2}. Compared with given methods, we can see MFAGAN achieves the best PSNR and LPIPS performance in most datasets. This reveals the effectiveness of our MFAM. Additionally, we also give the FLOPs, memory access cost, parameters, and latency for all the comparison methods. Our model achieves large compression ratios. In particular, our proposed method shows a clear advantage of ESRGAN compression compared to the previous GAN compression mthod AGD \cite{fu2020autogan}. We can reduce the memory access cost of the ESRGAN generator by 8.2$\times$, which is 2$\times$ better compared to the previous AGD method while achieving a much better PSNR and LPIPS. It demonstrates that MFAGAN is superior to other perceptual-driven methods in a comprehensive performance. 
\begin{table}[htp]\footnotesize
\setlength{\tabcolsep}{1.5pt}
\caption{Inference latency comparison of our MFAGAN with ESRGAN  on the Qualcomm Snapdragon 865 GPU  on 4$\times$ SR task.}
\vspace{-0.1in}
\newcommand{\tabincell}[2]{\begin{tabular}{@{}#1@{}}#2\end{tabular}}
\centering
\begin{tabular}{cccccc}
\toprule
Model&	\tabincell{c}{Memory\\(G)}	&\tabincell{c}{\#FLOPs\\(G)}&	\tabincell{c}{\#Param.\\(MB)}&\tabincell{c}{Set5\\PSNR$\uparrow$ LPIPS$\downarrow$}&\tabincell{c}{Mobile\\Latency}\\
\midrule
ESRGAN	&4.29&	291&	16.67&	\tabincell{c}{30.45 $\vert$ 0.0572}&1150\\	
\textbf{MFAGAN(ours)}&	0.52&	8.41&	0.55&\tabincell{c}{30.16 $\vert$ 0.0571}&70\\	
\bottomrule
\end{tabular}
\vspace{-8pt}
\label{t3}
\end{table}

\begin{figure}[]
\captionsetup[subfloat]{labelformat=empty,font=scriptsize}
\begin{minipage}[b]{0.37\linewidth}
\centerline{\subfloat[8023 from B100 (4$\times$)]{\includegraphics[width=1\linewidth]{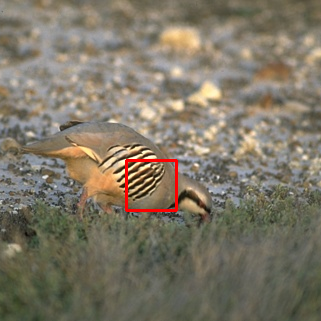}}}
\end{minipage}\hspace{0.05in}
\setcounter{subfigure}{0}
\begin{minipage}[b]{0.53\linewidth}
\centerline{
\subfloat[]{\includegraphics[width=0.28\linewidth]{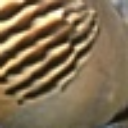}}\hspace{-2pt}\
\subfloat[]{\includegraphics[width=0.28\linewidth]{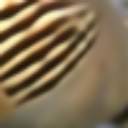}}\hspace{-2pt}\ 
\subfloat[]{\includegraphics[width=0.28\linewidth]{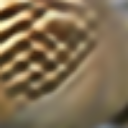}}\hspace{-2pt}\ 
\subfloat[]{\includegraphics[width=0.28\linewidth]{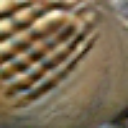}}}
\vspace{-0.15in}
\renewcommand\arraystretch{0.5}
\begin{tabular}{cccc}
\hspace{1pt}\scriptsize HR& \hspace{8pt}\scriptsize DRLN& \hspace{6pt}\scriptsize IMDN& \hspace{3pt}\scriptsize SRGAN\\
\end{tabular}\vspace{-0.15in}

\centerline{
\subfloat[]{\includegraphics[width=0.28\linewidth]{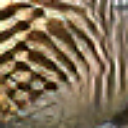}}\hspace{-2pt}\ \subfloat[]{\includegraphics[width=0.28\linewidth]{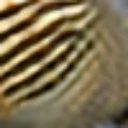}}\hspace{-2pt}\ \subfloat[]{\includegraphics[width=0.28\linewidth]{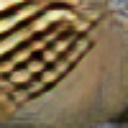}}\hspace{-2pt}\ \subfloat[]{\includegraphics[width=0.28\linewidth]{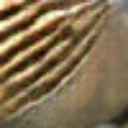}}}
\vspace{-0.15in}
\renewcommand\arraystretch{0.5}
\begin{tabular}{cccc}
\hspace{-12pt}\scriptsize EnhanceNet& \hspace{-8pt}\scriptsize ESRGAN& \hspace{3pt}\scriptsize AGD& \hspace{1pt}\scriptsize MFAGAN\\
\end{tabular}
\end{minipage}\vspace{-0.1in}\\
\begin{minipage}[b]{0.37\linewidth}
\centerline{\subfloat[223061 from B100 (4$\times$)]{\includegraphics[width=1\linewidth]{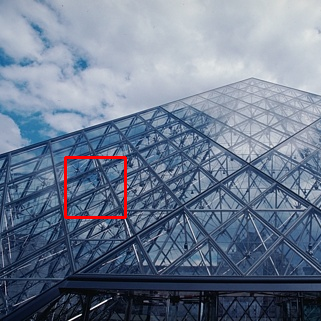}}}
\end{minipage}\hspace{0.04in}
\setcounter{subfigure}{0}
\begin{minipage}[b]{0.53\linewidth}
\centerline{
\subfloat[]{\includegraphics[width=0.28\linewidth]{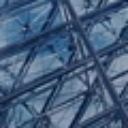}}\hspace{-2pt}\
\subfloat[]{\includegraphics[width=0.28\linewidth]{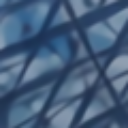}}\hspace{-2pt}\ \subfloat[]{\includegraphics[width=0.28\linewidth]{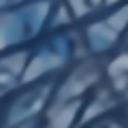}}\hspace{-2pt}\ \subfloat[]{\includegraphics[width=0.28\linewidth]{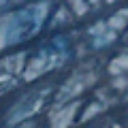}}}
\vspace{-0.15in}
\renewcommand\arraystretch{0.5}
\begin{tabular}{cccc}
\hspace{1pt}\scriptsize HR& \hspace{8pt}\scriptsize DRLN& \hspace{6pt}\scriptsize IMDN& \hspace{3pt}\scriptsize SRGAN\\
\end{tabular}\vspace{-0.15in}

\centerline{
\subfloat[]{\includegraphics[width=0.28\linewidth]{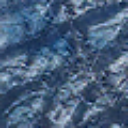}}\hspace{-2pt}\ \subfloat[]{\includegraphics[width=0.28\linewidth]{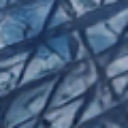}}\hspace{-2pt}\ \subfloat[]{\includegraphics[width=0.28\linewidth]{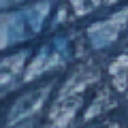}}\hspace{-2pt}\ \subfloat[]{\includegraphics[width=0.28\linewidth]{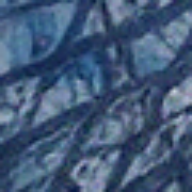}}}
\vspace{-0.15in}
\renewcommand\arraystretch{0.5}
\begin{tabular}{cccc}
\hspace{-12pt}\scriptsize EnhanceNet& \hspace{-8pt}\scriptsize ESRGAN& \hspace{3pt}\scriptsize AGD& \hspace{1pt}\scriptsize MFAGAN\\
\end{tabular}
\end{minipage}\vspace{-0.1in}\\
\begin{minipage}[b]{0.37\linewidth}
\centerline{\subfloat[223061 from B100 (4$\times$)]{\includegraphics[width=1\linewidth]{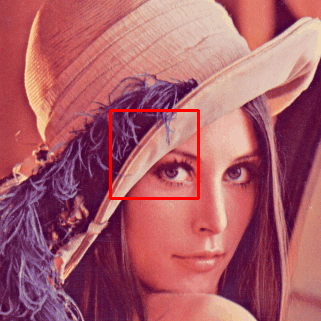}}}
\end{minipage}\hspace{0.04in}
\setcounter{subfigure}{0}
\begin{minipage}[b]{0.53\linewidth}
\centerline{
\subfloat[]{\includegraphics[width=0.28\linewidth]{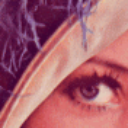}}\hspace{-2pt}\
\subfloat[]{\includegraphics[width=0.28\linewidth]{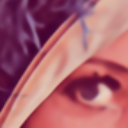}}\hspace{-2pt}\ \subfloat[]{\includegraphics[width=0.28\linewidth]{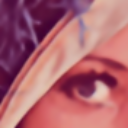}}\hspace{-2pt}\ \subfloat[]{\includegraphics[width=0.28\linewidth]{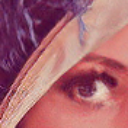}}}
\vspace{-0.15in}
\renewcommand\arraystretch{0.5}
\begin{tabular}{cccc}
\hspace{1pt}\scriptsize HR& \hspace{8pt}\scriptsize DRLN& \hspace{6pt}\scriptsize IMDN& \hspace{3pt}\scriptsize SRGAN\\
\end{tabular}\vspace{-0.15in}

\centerline{
\subfloat[]{\includegraphics[width=0.28\linewidth]{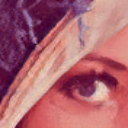}}\hspace{-2pt}\ \subfloat[]{\includegraphics[width=0.28\linewidth]{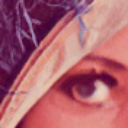}}\hspace{-2pt}\ \subfloat[]{\includegraphics[width=0.28\linewidth]{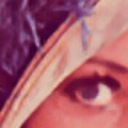}}\hspace{-2pt}\ \subfloat[]{\includegraphics[width=0.28\linewidth]{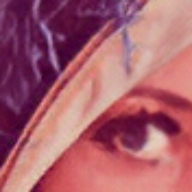}}}
\vspace{-0.15in}
\renewcommand\arraystretch{0.5}
\begin{tabular}{cccc}
\hspace{-12pt}\scriptsize EnhanceNet& \hspace{-8pt}\scriptsize ESRGAN& \hspace{3pt}\scriptsize AGD& \hspace{1pt}\scriptsize MFAGAN\\
\end{tabular}
\end{minipage}\vspace{-0.1in}\\
\caption{4$\times$ SR visual results for common test datasets. Our model (MFAGAN) can produce sharp edges and rich textures compared with other state-of-the-art methods. \textbf{(Zoom in for best view)}}
\vspace{-5pt}
\label{f9}
\end{figure}

\textbf{4$\times$  Qualitative comparisons.} To further illustrate the analyses above, we show visual comparisons on scales 4$\times$ on B100 and Set14. From Figure \ref{f9}, it can be observed that MFAGAN can generate sharp edges and realistic textures without introducing unpleasant artifacts. For image “8023” and “223061” in the B100 dataset, we can see that MFAGAN can recover the sharpness of the edges on the objects. Besides, for challenging details in the image “Lena” in Set14, MFAGAN can generate the correct textures of hair portion and hat edges. In general, MFAGAN achieves comparatively visual quality with ESRGAN and shows more realistic textures and sharper edges over IMDN and AGD.
\subsection{Comparison of Latency on Mobile Device}
Mobile inference acceleration has drawn people’s attention in recent years. At last, we compare the inference latency of our MFAGAN with ESRGAN on Qualcomm Snapdragon 865 GPU (on OPPO Find X2) for 4$\times$ SR images. We use the TensorFlow Lite framework to deploy the models on the mobile phone. The results are reported in Table \ref{t3}. We can see that our mobile-friendly model can process a  $128\times128$ input with 70 ms, while ESRGAN with comparable performance requires a significantly longer 1150ms. Our proposed MFAGAN achieves 16.4$\times$ measured speedup while saving memory usage by 88\%. The results show that MFAGAN is highly efficient in real-world applications. The algorithm and hardware co-design enables us to design specialized models on the target hardware.

\section{Conclusion}
In this work, we propose \textbf{M}ulti-scale \textbf{F}eature \textbf{A}ggregation Net based \textbf{GAN} (MFAGAN) compression framework to reduce the memory consumption of the generator in GAN-based SR. MFAGAN leverages memory-efficient architecture, layer-wise knowledge distillation, and hardware-aware evolutionary search to stabilize the training and improve the memory-efficiency. Extensive experiments show MFAGAN outperforms previous state-of-the-art methods with aggressively reduced memory access cost and a faster inference speed without visual quality degradation. For future works, we have plans to apply our findings to video SR.
{\small
\bibliographystyle{ieee_fullname}
\bibliography{references}
}
\end{document}